\documentclass[sigconf]{acmart}

\AtBeginDocument{%
  \providecommand\BibTeX{{%
    \normalfont B\kern-0.5em{\scshape i\kern-0.25em b}\kern-0.8em\TeX}}}

\setcopyright{acmcopyright}
\copyrightyear{2022}
\acmYear{2022}
\acmDOI{10.1145/1122445.1122456}

\acmConference[]{}{}{}
\acmBooktitle{}
\acmPrice{15.00}
\acmISBN{978-1-4503-XXXX-X/18/06}

\begin{document}

\title{Screen Correspondence: Mapping Interchangeable Elements between UIs}


\author{Jason Wu, Amanda Swearngin, Xiaoyi Zhang, Jeffrey Nichols, Jeffrey Bigham}
\email{{jsonwu,aswearngin,xiaoyiz,jwnichols,jbigham}@apple.com}
\renewcommand{\shortauthors}{Wu et al.}

\begin{abstract}
Understanding user interface (UI) functionality is a useful yet challenging task for both machines and people.
In this paper, we investigate a machine learning approach for \textit{screen correspondence}, which allows reasoning about UIs by mapping their elements onto previously encountered examples with known functionality and properties.
We describe and implement a model that incorporates element semantics, appearance, and text to support correspondence computation without requiring any labeled examples.
Through a comprehensive performance evaluation, we show that our approach improves upon baselines by incorporating multi-modal properties of UIs.
Finally, we show three example applications where screen correspondence facilitates better UI understanding for humans and machines: \textit{(i)} instructional overlay generation, \textit{(ii)} semantic UI element search, and \textit{(iii)} automated interface testing.
\end{abstract}



\begin{teaserfigure}
\includegraphics[width=\linewidth]{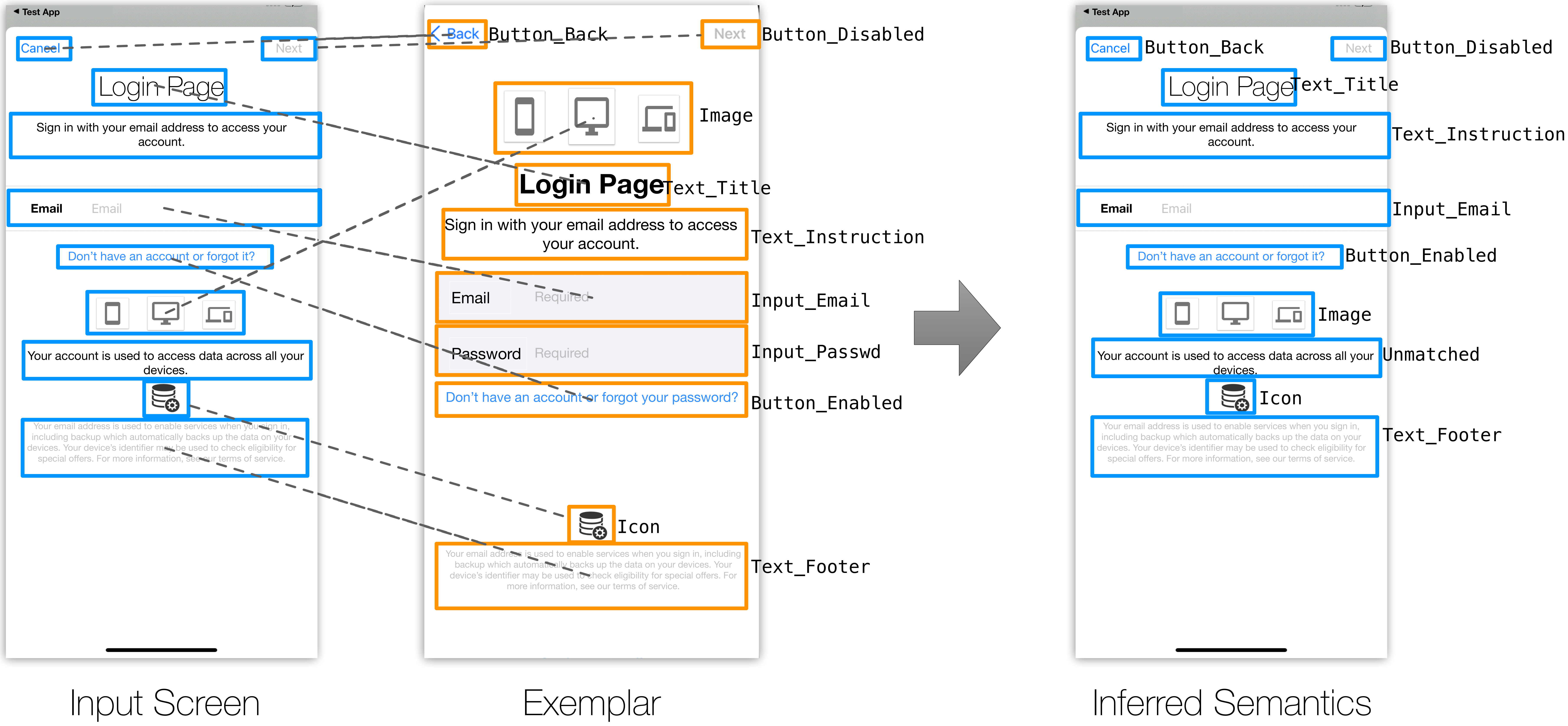}
  \caption{Screen correspondence produces a mapping of similar UI elements across two UIs that have related elements.
    Screenshots are encoded using a multi-modal model that segments and featurizes UI elements. Mappings are generated that link element pairs that have the same or similar functionality across UI screens.}
    \vspace{2pc}
  \Description{An input screenshot of the Mail app login screen is missing semantic information. An exemplar screenshot of the iCloud login screen contains labeled fields, which are transferred onto the Mail app.}
  \label{fig:teaser}
\end{teaserfigure}

\maketitle

\section{Introduction}
Understanding how user interfaces (UIs) can be operated to achieve some goal can be challenging for both machines and humans, especially those who are less tech-savvy.
While automated systems in the right circumstances can provide useful assistance \cite{yu2020maps,yeh2011creating} or automatically complete the task themselves \cite{li2017sugilite,li2020mapping}, people can be hindered or completely blocked by apps that do not provide necessary metadata, such as the view hierarchy.
A promising approach involves inferring UI functionality solely from the pixels rendered to the screen, but to date this method has primarily been useful for identifying the location and type of typical UI elements~\cite{zhang2021screen} and not higher-level semantics. For example, these algorithms can identify that a screen has a button that contains the text ``Login,'' but are unaware of the higher-level concept of logging in to a service, and they cannot infer what the role of this button would be in that process.


There are many higher-level semantics in user interfaces (e.g., login, account registration, shopping carts), which would correspond to an enormous number of classes if we attempted to use a classifier to predict their occurrence. Instead of making class predictions, an alternate approach to data inference involves directly comparing inputs to previously encountered examples with known properties.
Studies on human \cite{tomasello2005constructing} and machine \cite{vinyals2016matching} learning suggest that direct comparison is a useful tool, especially when relevant examples are available in the form of analogies \cite{chan2018solvent, hertzmann2001image} or templates \cite{guu2018generating,he2020learning}.
This concept can be highly effective for UIs, as many belong to the same app or are constructed to serve a similar purpose.
For example, knowledge of how an app screen was previously interacted with by an app crawler or automated UI tester could aid in producing more robust and consistent results when visited again.
Similar inferences can also be made for related screens from different apps, such as by determining that a button with the label ``Login'' in a new app is likely used to submit a login request because that is how a similar button is used in a known app.

In this paper, we pose the problem of \textit{screen correspondence} to map interchangeable elements between two UI screenshots (Figure \ref{fig:teaser}).
We introduce a multi-modal transformer model for detecting, featurizing, and matching UI elements.
Our approach is \textit{unsupervised}, which allows it to work without a large dataset of labeled examples, which could be costly and time-consuming to collect.
In a performance evaluation with strong baselines, we compare our approach to existing correspondence algorithms used in computer vision (CV) and heuristics such as schema-matching.
Our results indicate that our multi-modal model outperforms all existing baselines.

We describe and implement three example applications that show the utility of screen correspondence for humans and machines to understand UI functionality.
We create an application to \textbf{generate instructional overlays} by transferring high-quality human-authored coach marks (a type of instructional label) from one screen to another of the same category (\textit{e.g.,} two registration screens).
To support UI design search and exemplar-based exploration, we used our model to \textbf{index a large dataset of UI elements and screens}.
Finally, we built a system to \textbf{aid an automated app crawler} by identifying mappings between the elements of screens from different runs.

To summarize, we make the following contributions:
\begin{itemize}
    \item We introduce \textit{screen correspondence} as a method of mapping interchangeable elements between UI screens from their screenshots.
    \item We describe a machine learning approach to generating correspondence between two UI screenshots, and we show it outperforms existing baselines.
    \item We show the utility of screen correspondence in three example applications that improve both human and machine understanding of UI functionality.
\end{itemize}

\section{Related Work}
Our work is related to recent work in understanding user interfaces from their pixels, and also a variety of methods for understanding applications in terms of their many screens. We also overview machine learning solutions to correspondence problems in other domains, such as computer vision and natural language processing.

\subsection{Predicting Screen Semantics}
Computational representation of user interfaces are useful for many downstream tasks, such as design assistance \cite{leiva2020enrico, liu2018learning}, accessibility improvement \cite{zhang2021screen}, and task-oriented systems.
Screen Recognition \cite{zhang2021screen} generates accessibility metadata of a UI from screenshots using an object detection model and heuristics.
Screen Parsing \cite{wu2021screen} generates structured UI models from screenshots of UIs.
\textcolor{black}{Several models \cite{chen2020unblind,mehralian2021data,feng2021auto,zang2021multimodal} have also been trained to predict the semantics of unlabeled icons found in mobile apps. These models can be applied to improve the accessibility of mobile apps, either as a tool during design time or as an automated system that repairs existing apps at runtime.}
\textcolor{black}{Most of these models map UI elements to a pre-defined set of classes (\textit{e.g.} UI element and icon type), which may exclude less common components \cite{chen2022towards}.}

An alternative is to train models using self-supervision \cite{li2021screen2vec,fu2021understanding,deka2017rico}, which allows them to take advantage of larger unlabeled datasets.
Screen2Vec \cite{li2021screen2vec} and other pixel-based autoencoders \cite{deka2017rico,liu2018learning} map UIs to fixed-length embedding vectors which can be used to represent semantic properties.
The Pixel-words model \cite{fu2021understanding} employs a transformer model architecture and masked training objective based on prior work in NLP \cite{devlin2018bert}.
Our work builds upon these approaches to train a model for identifying UI element correspondences between screens.
\subsection{Multi-screen Understanding}
While many automated UI systems can benefit from understanding the semantics of a single screen, screens are rarely used in isolation.
Any task or interaction trace requires reasoning about multiple app screens and how they are related to each other.

StoryDroid is a system that extracts a storyboard of Android apps from APK files as an ``App Transition Graph'' \cite{chen2019storydroid}.
ActionBert \cite{he2020actionbert} models the relationship between two consecutive UI screens by predicting, among other things, which UI element was tapped on the first screen to reach the second (\textit{i.e.,} link component prediction).
\textcolor{black}{Longer sequences of touch interactions have also been modeled to better understand user behavior and app usage \cite{lee2018click,zhou2021large}.}

A particular problem that many multi-screen systems aim to address is identifying whether two screens are instances of the same UI, a problem which we refer to as ``screen fingerprinting.''
NEAR \cite{yandrapally2020near} detects near-duplicate pages on the web using a combination of visual and DOM-based features.
Prior work \cite{disfani2022understanding} used supervised learning to predict the relationship and transitions between screenshots by, among other things, classifying whether inputs were different instances of the same screen (\textit{e.g.,} a news app with dynamic content).

Screen fingerprinting is useful for comparing screens to known examples; however, finer grain mappings (\textit{e.g.,} element-level fingerprinting) can result in higher fidelity comparisons and additional benefits.
Bricolage \cite{kumar2011bricolage} is a system that renders the content of one web page using the style and layout of another. It employs a supervised element matching model that featurizes web elements based on their DOM representation and was trained on a dataset of human-generated mappings.
Interaction proxies \cite{zhang2018robust} rely on a set of equivalency heuristics to identify UI components and structures found in Android view hierarchies to facilitate accessibility repair.

Our work is related to these approaches in multi-screen understanding and specifically element fingerprinting.
While many previous examples relied heavily on the availability of a structured UI representation (\textit{e.g.,} DOM, view hierarchy) and were trained on labeled data, our approach requires only screenshots of related apps with optional labels.
\subsection{Machine Learning of Correspondence}
Machine learning has been used to learn correspondences in other domains, such
as computer vision (CV) and natural language processing (NLP), which are closely related to our approach.

A longstanding problem in CV is inferring accurate correspondence of objects from different images.
Homography estimation \cite{hartley2003multiple} involves finding a mapping, either sparse or dense, or a transformation matrix that describes perspective changes in two images of the same object or scene.
Optical flow \cite{horn1981determining} applies a similar concept to finding mappings between consecutively taken images.
A common approach involves computing appearance descriptors (keypoints), then creating a mapping that optimizes the global correspondence \cite{fischler1981random, liu2010sift}.
Recent work \cite{aberman2018neural} has extended these approaches using learned semantic features to infer correspondence between images of inter-class or inter-domain objects.


Correspondence learning has also been useful for many tasks in NLP such as pronoun co-reference resolution and commonsense reasoning, both of which rely on modeling correspondences between words to resolve ambiguities \cite{wu2019scratchthat,Kocijan2019ASR,levesque2012winograd}.
Language translation and understanding, particularly for low-resource languages, benefit from learning word alignments to higher-resource languages \cite{dyer2013simple,sabet2020simalign}.
Finally, other types of conditional natural language generation have benefited from learning alignments between words with similar meanings \cite{bahdanau2014neural,see2017get}.

Screen correspondence is related to many machine-learning driven approaches to identifying correspondence.
Our transformer model builds upon many of these approaches by combining visual information and word-alignment techniques to produce screen correspondence.
In our evaluation, we compared our system to several baseline techniques from the CV and NLP literature.
We show that by incorporating multiple sources of information, our model generates better representations for UI elements, which leads to more accurate correspondence predictions.

\section{Screen Correspondence}
We define \textit{screen correspondence} as the task of mapping interchangeable UI elements between two UI screens.
While matching UI elements between screens may seem simple, it is a complex problem (especially from pixels alone) with many practical use-cases.
Previous work relied on mappings to retarget UIs \cite{kumar2011bricolage}, provide help \cite{yeh2011creating,yu2020maps}, assist design \cite{bunian2021vins}, and test GUIs \cite{chang2010gui} and specifically called for more robust matching to improve performance. 


We primarily consider cases where two UI screens are of the same category (\textit{e.g.,} Login or Registration) but from different apps (\textit{i.e.,} intra-class examples).
This is challenging because UI element pairs across such screen type pairs may not share similar appearance, text, or position.
Instead, the model must reason about the purpose of each element in the context of its screen.
To give intuition why even a seemingly simple example is hard, consider two Login screens (Figure~\ref{fig:teaser}): one contains \textit{Username} and \textit{Password} fields, while another contains \textit{Email} and \textit{Password} fields. Position and element type information alone is unreliable for matching, since the text fields may have different sizes or appear at different locations on each screen.
Appearance information alone is also noisy for matching, since the text fields may have different visual themes.
State-of-the-art text encoders, even those trained on phrases, are unreliable (\textit{e.g.,} most text models would produce a higher similarity score for \textit{Username} and \textit{Password} than \textit{Email}).

\begin{figure*}
    \centering
    \includegraphics[width=0.8\textwidth]{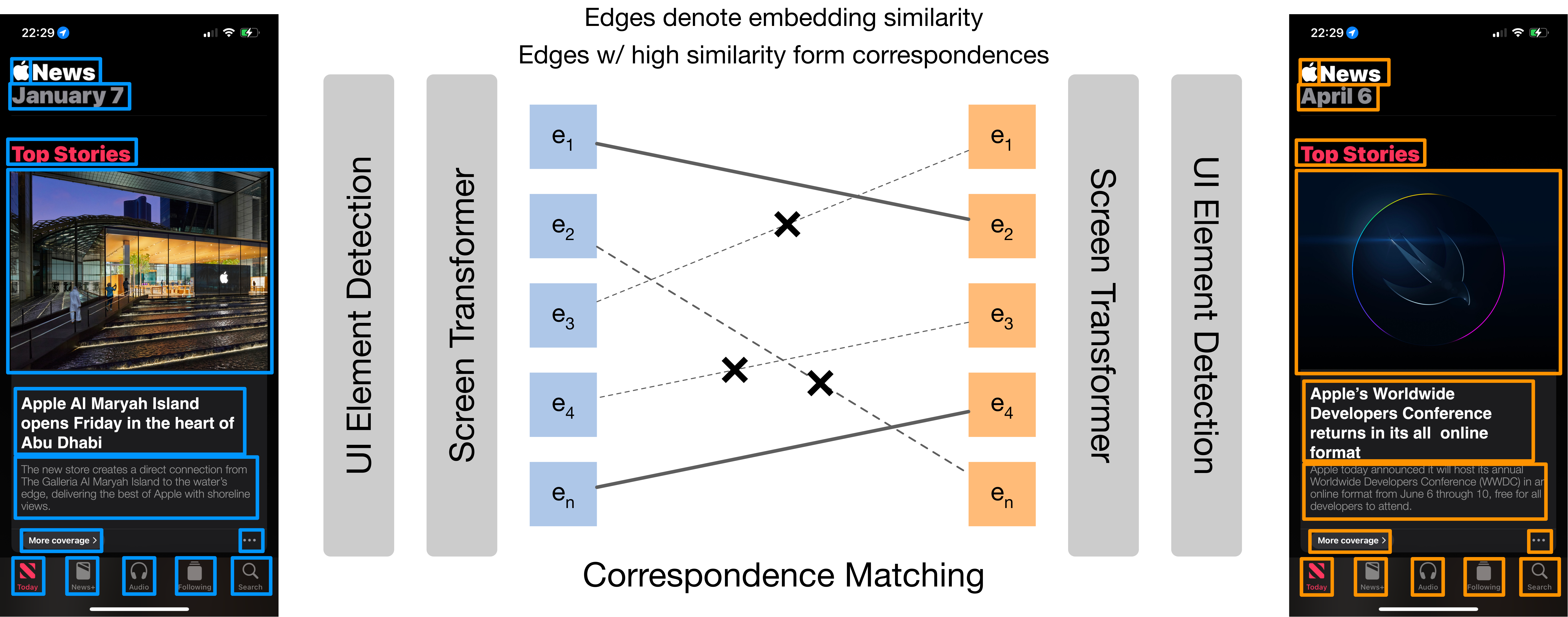}
    \caption{Overview of our screen correspondence approach. Elements and text from two screenshots are first extracted using UI element detection then featurized using a screen transformer model. Finally, a correspondence between UI elements are generated from element pairs with highly similar embeddings relative to other candidates.}
    \label{fig:correspondence_diagram}
\end{figure*}

To detect UI element correspondences between different UI screens, we built a system that \textit{(i)} automatically detects UI elements and text from screenshots, \textit{(ii)} generates multi-modal embeddings for each element, and \textit{(iii)} establishes mappings between individual UI elements with high similarity.
Figure \ref{fig:correspondence_diagram} shows a high-level overview of our approach.
\subsection{UI Element Detection} 
The first stage of our system identifies semantically relevant pieces of information from a UI screenshot, such as UI elements and text.
The input is a bitmap and the output is a list of detected UI elements and pieces of text.

We use a pre-trained object detection model from previous work \cite{wu2021screen} that was trained to recognize UI elements in iOS app screens.
The pre-trained model uses the Faster-RCNN architecture \cite{ren2015faster} and was trained on the \textsc{AMP} dataset \cite{zhang2021screen}, which is separate from the main dataset used in this paper.
It achieved a class-weighted mAP score of 0.8.
We use post-processing procedures, such as score-based thresholding and inter-class non-max suppression (NMS), to improve the quality of the output.
Optical character recognition (OCR) is performed using Tesseract \cite{smith2007overview}, an open-source, off-the-shelf OCR software package.
We run OCR on regions of the screen that correspond to text elements as detected by our element detector.

\subsection{UI Element Encoder}
Using the elements segmented from the screenshot, we generate representations that encode properties useful for comparison.
In our work, we consider relative positioning, element/icon category, visual appearance, and text, properties which we hypothesize to be relevant to element semantics.
\textcolor{black}{We used pre-trained models to predict these properties (element detection \cite{wu2021screen} and icon type (``common icon classifier" from previous work \cite{chen2022towards})) from screenshots. Note that the pre-trained models were trained on different datasets (\textit{i.e.,} no sample overlap) than the ones used in our paper.}

\begin{figure*}
    \centering
    \includegraphics[width=0.85\textwidth]{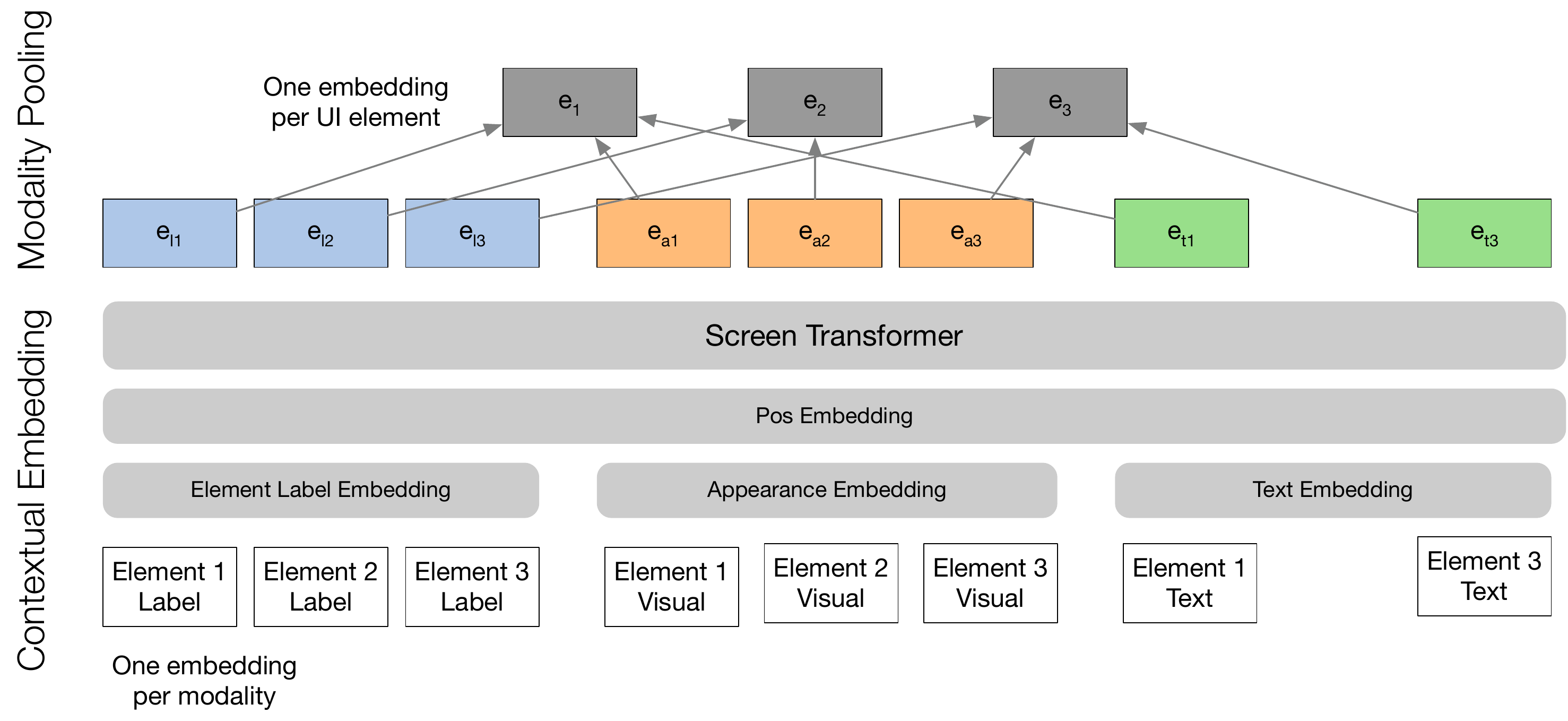}
    \caption{Architecture diagram of our screen transformer model. Each modality-specific input is treated as separate inputs to our transformer model, which implicitly aligns them based on their positional information. Note that elements may be missing modalities (Element 2 in this example). After the per-modality inputs are processed by our transformer, we generate element embeddings (\textit{i.e.,} one per element) by pooling together outputs corresponding to the same original element.}
    \label{fig:model_architecture_diagram}
\end{figure*}
\subsubsection{Modality Representations}
We use off-the-shelf models to generate modality-specific features for each element, then feed their output into a screen transformer model, which combines and learns further associations between them.

\textit{Positional information:}
Previous work \cite{disfani2022understanding,fu2021understanding} encoded element position as a simple concatenation of bounding box coordinates.
We hypothesized that relative position may be more effective, since UI interactions such as scrolling, text flow, and dynamic content could cause changes in absolute position but have less effect on relative ordering.
We adopted a relative positional encoding scheme used to improve the performance of language models \cite{shaw2018self} that incorporates pairwise distance when calculating the attention score between two elements.

\textit{Element Category:}
We categorized elements based on their UI and icon type.
Our pre-trained element detector classifies elements into 12 categories, as defined by previous work \cite{zhang2021screen}.
Three of these can be further delineated into sub-categories.
We separate the Toggle and Checkbox classes based on their selection state (\textit{e.g.,} Toggle on and Toggle off).
We also classified common icon types using a separate pre-trained CNN model \cite{zhang2021screen}.
In total, we consider 83 unique categories of elements and represent them as one-hot vectors.

\textit{Visual Appearance:}
\textcolor{black}{We featurized regions using the intermediate representations of a proposal-based object detector. Similar approaches have also been used by visual question-answering models, which also need to take into account multiple visual information \cite{singh2019towards.}}
Since our UI element detector is based off of a similar proposal-based architecture, we retrieve the activations of the object proposals corresponding to detected elements using the \texttt{fc6} layer \cite{singh2019towards}.
This approach to featurizing appearance is beneficial since it results in a fixed-size representation for image regions without the need to explicitly resize or crop them.

\textit{Text:}
Numerous embedding methods have been developed for representing words, sentences, and documents.
Sentence transformers are transformer-based models for encoding variable-length text into an embedding space representative of semantic meaning \cite{reimers2019sentence}.
Since much of the text on UI screens is relatively short, we use a variant specifically trained to encode phrases \cite{wang2021phrase}.

\subsubsection{Transformer Model}
To further enrich and learn associations between the modality-specific element representations, we designed a model that generates a fixed-size embedding for each detected UI element.
Our model is based on the transformer architecture (Figure \ref{fig:model_architecture_diagram}), which has been used for UI representation learning \cite{he2020actionbert,fu2021understanding}.
The modifications we described (\textit{e.g.,} relative positioning and appearance features) are aimed at improving performance on the correspondence task.

Because not all elements have the same attributes, \textit{e.g.,} not all UI elements have text, we rely on the transformer's attention mechanism to implicitly align information.
Each modality-specific representation with the exception of position is first embedded with a separate linear layer to a common size.
Instead of creating one input vector for each element by concatenating features from each modality, we create an input vector for each modality of each element.
For example, a login button could result in three input vectors for element category, visual appearance, and text.
All inputs are fed into a series of stacked self-attention blocks, which results in one output embedding for each original input vector.
Finally, we use pooling to recover the output embeddings associated with each original UI element and compute their mean to incorporate information from all of the modalities.

\subsubsection{Unsupervised Training}
We did not have access to labeled data during the development of our model, so we used unsupervised training to learn its parameters.
Masked element prediction is a training objective that requires the model to reconstruct an input that has been corrupted by randomized masking (\textit{i.e.,} replacing a portion of the input with 0's).
Previous work \cite{tan2019lxmert} has shown that this training objective encourages the model to learn semantically relevant representations since it must learn to associate masked information with other sources of information.

The reconstruction loss was measured separately for all modalities (element category, visual appearance, and text) then added together to obtain the model's total loss.
L2-loss was used for reconstruction of visual and text features, and cross-entropy loss was used for reconstruction of element category.

\subsection{Correspondence Matching}
After we used our screen transformer model to featurize UI elements on two screens, we perform a matching procedure to predict correspondences between them.
For a pair consisting of a source screen with $M$ elements and a target screen with $N$, we construct a $M \times N$ cost matrix $C \in \mathbb{R}^{M x N}$ to represent correspondence scores.
The matching cost $C_{i, j}$ is computed using cosine similarity.


\textcolor{black}{Several approaches have been used to generate correspondences from cost matrices \cite{sabet2020simalign}. A simple approach of matching based solely on highest cosine similarity may make suboptimal decisions when one element has more than one likely match.}
\textcolor{black}{In our final implementation, we} formulated correspondence mapping as an optimization problem that finds the assignment between two sets that results in the lowest overall cost \cite{kuhn1955hungarian}.
We employ this approach for matching elements between screens, since elements are more likely to be dissimilar.
To reduce false positives, we employ additional pre-processing and post-processing steps.
Before running the best-cost optimization, we prune unlikely matches from the cost-matrix so that each element only considers its $k$ closest neighbors.
Afterwards, we ignore matches where $C_{i, j} < c$. We tuned the values of $k=5, c=0.4$ based on manual examination of a small number of examples.
\textcolor{black}{This approach is similar to approaches in text decoding models that consider only the top $k$ most likely tokens, which have been shown to generate higher quality output by reducing the effect from low-probability outputs.}
\section{Dataset}
\textcolor{black}{We developed and trained our transfer model on two datasets of app screens that were generated by manual crawling of popular mobile apps: \textsc{Crawls} and \textsc{Rico}.}
\textcolor{black}{The \textsc{Crawls} dataset, which was used by prior UI modeling work \cite{chen2022towards}, consists of 750,000 iOS app screens from 6,000 apps and was collected by crowdworkers who were instructed to manually explore mobile applications through a remote interface that periodically captured screenshots and additional metadata of the current app screen.}
\textcolor{black}{The \textsc{Rico} dataset \cite{deka2017rico} is a publicly available dataset of 72,000 Android screens from 9,700 apps that was also collected by crowdworkers remotely operating devices.}
We divided \textcolor{black}{each} dataset into training (70\%), validation (15\%), and testing (15\%) splits by their crawl ID, which corresponds to which app was crawled.

\subsection{Evaluation Dataset}
While our training algorithm does not depend on labeled data (\textit{i.e.,} unsupervised), we manually collected a small set of labeled examples (\textcolor{black}{900 pairs across 90 screens}) \textcolor{black}{from each dataset} to evaluate our system.

\subsubsection{Data Collection}
Our evaluation dataset consists of data from 9 types of screens that we hypothesized could have correspondences: Media Player, In-App Purchases, Login, Permission Request, Register, Pre-Login, Pop-up, Search, and Web Views.
We initially asked crowdworkers to categorize a set of screenshots outside of the training split based on a criteria for each category.
\textcolor{black}{Unlike \textit{app categories}, which might be used to categorize apps (\textit{e.g.,} finance, health, social media), we focused on \textit{screen categories}, since both a health app and a banking app might both contain a login screen that could contain correspondences.}
\textcolor{black}{For each of our two datasets, we} sampled a small number of screens from each category for correspondence labeling \textcolor{black}{(9 categories x 10 screens = 90 screens total).}
The 9 categories that we chose do not cover all possibilities, but we believe they constitute a reasonable subset.
More detailed descriptions and criteria of each category is available in the appendix of this paper.

\subsubsection{Data Labeling}
%
We created a labeling interface to annotate our small evaluation split.
First, a randomly selected element was shown on a screen, and the interface displayed a prompt asking if the element was likely to appear on other screens of the same type: \textit{``Are elements of similar functionality likely to appear on other Login screens?''}
If the user responded ``Yes,'' the application displayed a prompt for a label: \textit{``What is the role of this element in the current screen?''}
We built our interface to include auto-complete functionality to encourage labelers to identify correspondence categories that could generalize across screens, \textit{e.g.,} ``login button'' instead of ``button to log into my credit card account.''
The autocomplete list was pre-populated with 5 choices for each category and was auto-updated with novel descriptions.
If a label was provided on the first step, then the user was shown other screens from the same category and asked to select elements with a similar role, if they were present on the screens.

A drawback of this approach is that it is slow, since it requires providing a role description before elements are matched.
However, we found the additional consideration of element role is useful for reasoning about correspondences.
\section{Evaluation}
We evaluate our model against several baselines and ablated versions of our model.
Our results show that compared to heuristic and traditional key-point methods used in CV, multi-modal transformer encodings lead to better correspondences.
Furthermore, our ablation experiments show that the architectural improvements we made lead to modest performance gains.

\subsection{Baselines}
In this section, we describe the baselines used in our performance evaluation.
We focus primarily on other \textit{unsupervised} approaches, since our constraint was that we didn't have any labeled data available for training.
Similar supervised approaches exist \cite{kumar2011bricolage}, but they depend on element-level annotations and access to underlying source code (\textit{i.e.,} HTML).

For comparison, we chose a variety of baselines that include keypoint-based methods used for image matching and heuristics such as schema-matching.
Our main constraint was that we did not have large quantities of labeled data for supervised machine learning methods, so we selected unsupervised techniques for comparison.



\textit{ORB:}
As a review, image correspondence relies on the computation of semantic features from regions of the image.
Semantic features, usually invariant to surface-level changes such as translation and scale, are first calculated for small, localized regions of the image.
When this process is repeated recursively, the receptive field increases, and globally-aware features can be learned.
ORB \cite{rublee2011orb} is a traditional CV approach to generating descriptor features.
We first computed ORB descriptors for each screenshot which resulted in numerous keypoints at salient points of the image.
Using brute-force matching, keypoints from one image were matched onto keypoints from another image based on descriptor similarity. 
Finally, to translate keypoint similarity to UI element similarity, we used an object detector to compute the boundaries of UI elements and matched elements based on the number of matching keypoints contained within them.

\textit{Neural Best Buddies:}
Neural Best Buddies (NBB) uses the internal representations of a deep CNN to featurize and match image regions.
Like ORB, it also generates keypoint descriptors but uses activations from a convolutional neural network (CNN).
One advantage that CNN features have over traditional methods is that learned features can better correspond to semantic properties that the network was trained on (e.g., image classification).
To run our experiments, we used the code released by the authors of the paper~\footnote{\url{https://github.com/kfiraberman/neural_best_buddies}}.
The original paper focuses on finding correspondences between ``natural images'' and use a VGG-19 model \cite{simonyan2014very} that was pretrained on ImageNet~\cite{deng2009imagenet}.
Since UI screenshots have different properties than images found in ImageNet, we initially tried to train a CNN model better representative of UIs using an unsupervised autoencoder objective due to the lack of labels in our training set.
However, we found the autoencoder model did not produce good outputs, so we report results using the pre-trained ImageNet model.

\textit{Schema-matching Heuristic:}
One drawback of keypoint-based methods that we explored is that keypoints are generated using the entire image as input and without knowledge of UI element locations.
Schema-matching is an approach that first considers each predicted element as a discrete object, then uses its attributes (\textit{i.e.,} schema) to compare similarity to other candidates.
We implemented a heuristic that uses schema matching through incorporating the predicted UI element/icon type by concatenating their one-hot class predictions into a single vector and applying the same best-cost matching algorithm \cite{kuhn1955hungarian}.
More sophisticated schema-matching may incorporate additional information, such as UI hierarchy (\textit{e.g.,} an element that belongs in a list should be matched to another element in a list).
While possible to predict \cite{wu2021screen}, we did not incorporate hierarchical information since it requires complex techniques for tree matching but expect it would perform similarly to \cite{kumar2011bricolage}, which uses hierarchical information.

\textit{Screen Transformer Ablations:}
\textcolor{black}{Our} performance evaluation includes ablated variations of our main transformer model.
Transformer models allow learning more sophisticated representations of elements through data, which provides advantages over manually-defined schemas.
We evaluate several ablated versions of our model to understand the performance impact of our architectural changes.
Specifically, the ablated versions of our transformers removes certain components that we hypothesized to improve correspondence matching, such as relative positional embedding, visual appearance information, and text.
In addition, we evaluated the Pixel-words transformer \cite{fu2021understanding}, which our model is based on, but we adjust the number of element classes, layers, attention heads, and hidden dimensions to be the same as our other models.
The Pixel-words transformer also includes a ``layout embedding" network which featurizes the layout of UI using a semantic map which is fed into an autoencoder.
To summarize, the Pixel-words configuration \textit{(i)} considers categorical and text information, \textit{(ii)} uses absolute positional encodings, and \textit{(iii)} includes an additional layout embedding component.


\subsection{Results}
\begin{table}[]
\small
\centering
\caption{\textcolor{black}{Performance of our approach and other baselines for screen correspondence. Our approach leads to the best performance, reaching an F1 score 0.61. We also included ablated versions of our model without relative positional embeddings, appearance features, and text features.}}
\begin{tabular}{@{}lllllll@{}}
\toprule
                                    & \multicolumn{3}{l}{CRAWLS} & \multicolumn{3}{l}{\textcolor{black}{RICO}} \\ \midrule
Model Configuration                 & P       & R       & F1     & P      & R      & F1     \\ \midrule
Screen Trans.                 & 0.66    & 0.57    & 0.61   & \textcolor{black}{0.83}      & \textcolor{black}{0.41}      & \textcolor{black}{0.55}      \\
Screen Tran. (w/o Relative)   & 0.58    & 0.53    & 0.56   & \textcolor{black}{0.74}      & \textcolor{black}{0.37}      & \textcolor{black}{0.49}      \\
Screen Trans. (w/o Appearance) & 0.74    & 0.44    & 0.55   & \textcolor{black}{0.77}      & \textcolor{black}{0.38}      & \textcolor{black}{0.51}      \\
Screen Trans. (w/o Text)       & 0.66    & 0.63    & 0.59   & \textcolor{black}{0.77}      & \textcolor{black}{0.37}      & \textcolor{black}{0.50}      \\
Screen Trans (Pixel-words)    & 0.70    & 0.49    & 0.58   & \textcolor{black}{0.83}      & \textcolor{black}{0.22}      & \textcolor{black}{0.35}     \\
Heuristic                           & 0.48    & 0.59    & 0.53   & \textcolor{black}{0.80}      & \textcolor{black}{0.32}      & \textcolor{black}{0.45}      \\
ORB                                 & 0.25    & 0.17    & 0.20   & \textcolor{black}{0.63}      & \textcolor{black}{0.21}      & \textcolor{black}{0.31}      \\
NBB                                 & 0.22    & 0.15    & 0.18   & \textcolor{black}{0.58}      & \textcolor{black}{0.17}      & \textcolor{black}{0.26}     \\ \bottomrule
\end{tabular}
\label{tab:results}
\end{table}
\begin{figure}
    \centering
    \includegraphics[width=20pc]{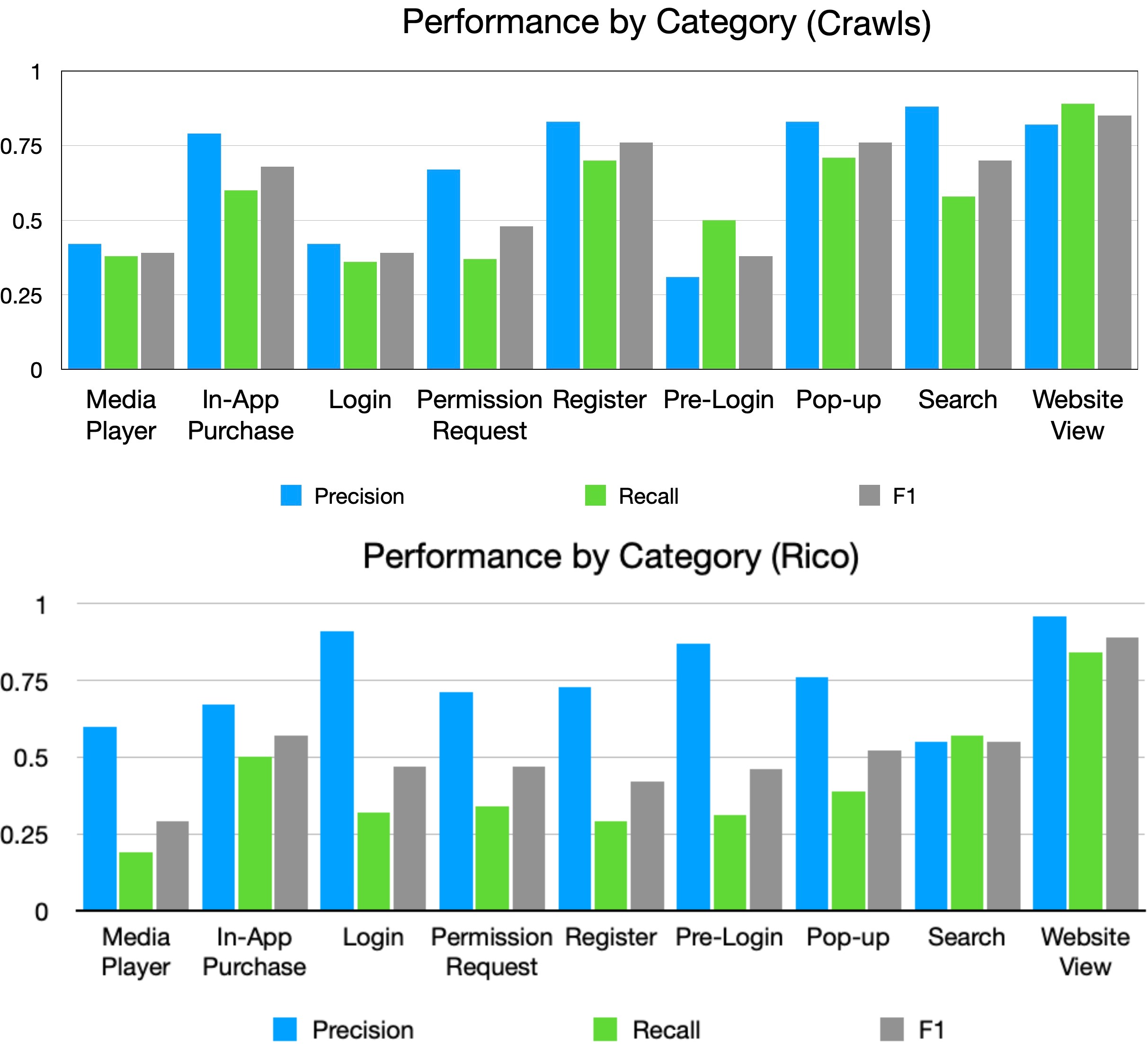}
    \caption{
    \textcolor{black}{Performance across different categories in the \textsc{Crawls} (Top) and \textsc{Rico} (bottom) datasets using the full Screen Transformer model configuration. The average classification performance was F1=0.61 on \textsc{Crawls} and F1=0.55 on \textsc{Rico}.}}
    \label{fig:category_perf}
\end{figure}

\subsubsection{Baseline Comparison}
Our evaluation results (Table \ref{tab:results}) shows the benefit of our multi-modal model over simpler baselines.
We employ standard classification metrics to measure the accuracy of element-to-element correspondences generated by our model.
Since elements in our evaluation dataset are labeled using their ground truth bounding boxes instead of our element detector's predictions, we first match predicted detections to ground truth elements using the best Intersection-over-Union (IoU) score.
Due to our labeling procedure where one element is highlighted at a time, the examples in our evaluation set were only partially labeled, meaning that screens contained only a randomly sampled subset of all possible corresponding pairs.
Our best model configuration reaches an F1 score of 0.61.
Screens in our dataset contained an average of around 20 elements, so correct correspondence required finding the best out of match out of all possible candidates.
The \textit{ORB} and \textit{NBB} baselines are based on keypoint-based matching, which is commonly used in CV to compare images.
Among them, ORB performs the best, achieving higher correspondence accuracy but performed poorly due to low recall.
One possible reason is that keypoints are generated at visually salient locations of the image, such as edges and corners, and without any knowledge of where UI elements are.
Thus, some UI elements may not contain many keypoints within them, reducing the quality of matches.
The schema-matching heuristic performed substantially better than keypoint-based methods and reached high recall by directly using the outputs of pre-existing models (\textit{i.e.,} element detection, icon classification).
Precision was lower, possibly due to the difficulty of accurately matching ambiguous elements without knowledge of additional context.

Our ablation experiments revealed that our modifications to the base transformer architecture led to modest improvements in terms of F1 score but also had other consequences for precision and recall.
For example, our model trained without appearance information was the lowest performing variation but reached the highest precision score.
We attribute these variations to the information encoded in each modality and may warrant different configurations based on intended use-case.


\subsubsection{Performance across \textcolor{black}{UI} Categories}
Figure \ref{fig:category_perf} provides a more in-depth breakdown correspondence by UI category.
\textcolor{black}{Our model achieved the best performance on the \textit{Website View} and \textit{In-App Purchase} categories and the worst performance on the \textit{Media Player} and \textit{Pre-Login} categories.}

One major source of error for our model was the presence of sub-categories within our dataset.
For example, we manually examined examples from the \textit{Pre-Login} and \textit{Login} categories, which received relatively low performance.
We discovered a considerable difference between apps that used different authentication providers, such as OAuth and Single-Sign-On (SSO).
For example, a ``traditional'' login screen might include text fields for entering a username and password, but an app using a SSO provide (\textit{e.g.,} Sign in with Apple) might only contain a button without any text fields.
We found that there was also variance within media player screens -- video players and music players had significant differences and some media players were full screen while others were not.
Since our correspondence model uses contextual information (\textit{i.e.,} information from other elements on the same screen) and relative positional encoding, this could significantly affect the computed representation.
One strategy to address this is the formation of sub-categories with a more consistent set of elements \textit{e.g.,} creating separate categories for traditional login screens and those with other types of authentication.
\textcolor{black}{\subsubsection{Performance across Datasets}
We evaluated all models and baselines on both the \textsc{Crawls} and \textsc{Rico} dataset. 
Overall performance between the two datasets were similar, although the \textsc{Rico} models performed slightly worse (F1=0.55) than ones trained on \textsc{Crawls} (F1=0.61). One possible reason for the performance discrepancy is that \textsc{Crawls} is an order of magnitude larger and the model was exposed to more variation during training time, which is beneficial for unsupervised training techniques. While the full transformer model is the best-performing configuration for both datasets, the relative performance ablated models were affected differently. Notably, the \textsc{Rico} models without text and the Pixel-Words model performed much worse, suggesting that its evaluation set may have contained more text-heavy screens.
}
\subsubsection{Correspondence between Same-screen Pairs}
\begin{table}[]
\centering
\caption{Performance of our approach and other baselines screen correspondence for \textit{same-screen} pairs \textcolor{black}{in the \textsc{Crawls} dataset}. Many configurations, including our model, reach a maximum F1 score of 0.76. We attribute labeling noise and the IoU element matching process used to assign predicted element locations to ground-truth boxes.}
\label{tab:same_screen_table}
\begin{tabular}{llll}
\hline
  Model Configuration                                     & P    & R    & F1   \\ \hline
Screen Transformer                  & 0.85 & 0.68 & 0.76 \\
Screen Transformer (w/o Relative)       & 0.86 & 0.68 & 0.76 \\
Screen Transformer (w/o Appearance) & 0.87 & 0.66 & 0.75 \\
Screen Transformer (w/o Text)       & 0.85 & 0.68 & 0.76 \\
Screen Transformer (Pixel-words)                            & 0.88 & 0.66 & 0.76 \\
Heuristic                              & 0.87 & 0.66 & 0.75 \\
ORB                                    & 0.78 & 0.48 & 0.59 \\
NBB                                    & 0.53 & 0.25 & 0.34 \\ \hline
\end{tabular}
\end{table}
In addition to evaluating our models on screens from different related apps (\textit{i.e.,} \textit{intra-class} pairs), we also investigated performance on \textit{same-screen} pairs.
Same-screen correspondence is useful for identifying the same UI element across multiple versions of the same screen.
For example, an app's appearance may change following an update or from dynamically updated content (\textit{e.g.,} a news page loads content from a remote source).
Following prior work \cite{disfani2022understanding}, we consider two screenshots to be the ``same'' if they represent different instances of the same underlying implementation, possibly with significantly different appearance.
Correspondence mapping can help guide automated systems such as crawlers to behave more consistently in these situations.
We randomly selected screen groups with the same app ID as those in the testing split of our \texttt{Crawls} dataset, then randomly sampled two screenshots from each group, resulting in 888 total pairs.
Upon manual inspection, we found that some of the sampled pairs had only minimal visual changes.
To filter out ``easy pairs,'' we constructed a heuristic that attempted to match elements based only on bounding box location.
If all elements in a pair were successfully matched, we discarded the example, since it meant that no significant dynamic change (\textit{e.g.,} scrolling, dynamic content) occurred.
After this process, the final dataset contains 607 examples.
\textcolor{black}{We did not repeat this for the \textsc{Rico} dataset because the authors applied a heuristic to filter out repeated views of the same screen \cite{deka2017rico}.}

Our observations and performance results (Table \ref{tab:same_screen_table}) show that \textit{same-screen} correspondence is generally higher.
Since \textit{same-screen} pairs are usually more visually similar, the model can rely more heavily on surface-level features and in many cases perform direct matching, such as looking for recurring text.
Many configurations, including our model, reached a maximum F1 score of 0.76.
Errors from labeling noise and IoU element matching (\textit{e.g.,} matching ground truth bounding boxes to predictions) may have established an effective ceiling, since our element detection model introduced errors (has a class-weighted mAP score of 0.8).



\section{Example Applications}
We describe three example applications that show the utility of screen correspondence to human and machine understanding of UI functionality.
Generating and transferring a type of instructional overlay called coach marks can help users navigate unfamiliar UIs by mapping them to previously encountered ones of the same class.
UI search is useful for app designers to find how concepts are expressed across apps (\textit{e.g.,} what are different ways of expressing a search intent?).
Finally, automated GUI testing can be made more robust by accounting for variations in visual presentation between different app versions without requiring platform-specific APIs or metadata.
These example applications are not meant to be novel, but we believe they show that accurate screen correspondence allows many existing systems to work under a wider range of conditions, \textit{e.g.,} using pixel data alone or improved robustness to dynamic visual changes.
\subsection{Instructional Overlays}
\begin{figure}
    \centering
    \includegraphics[width=15pc]{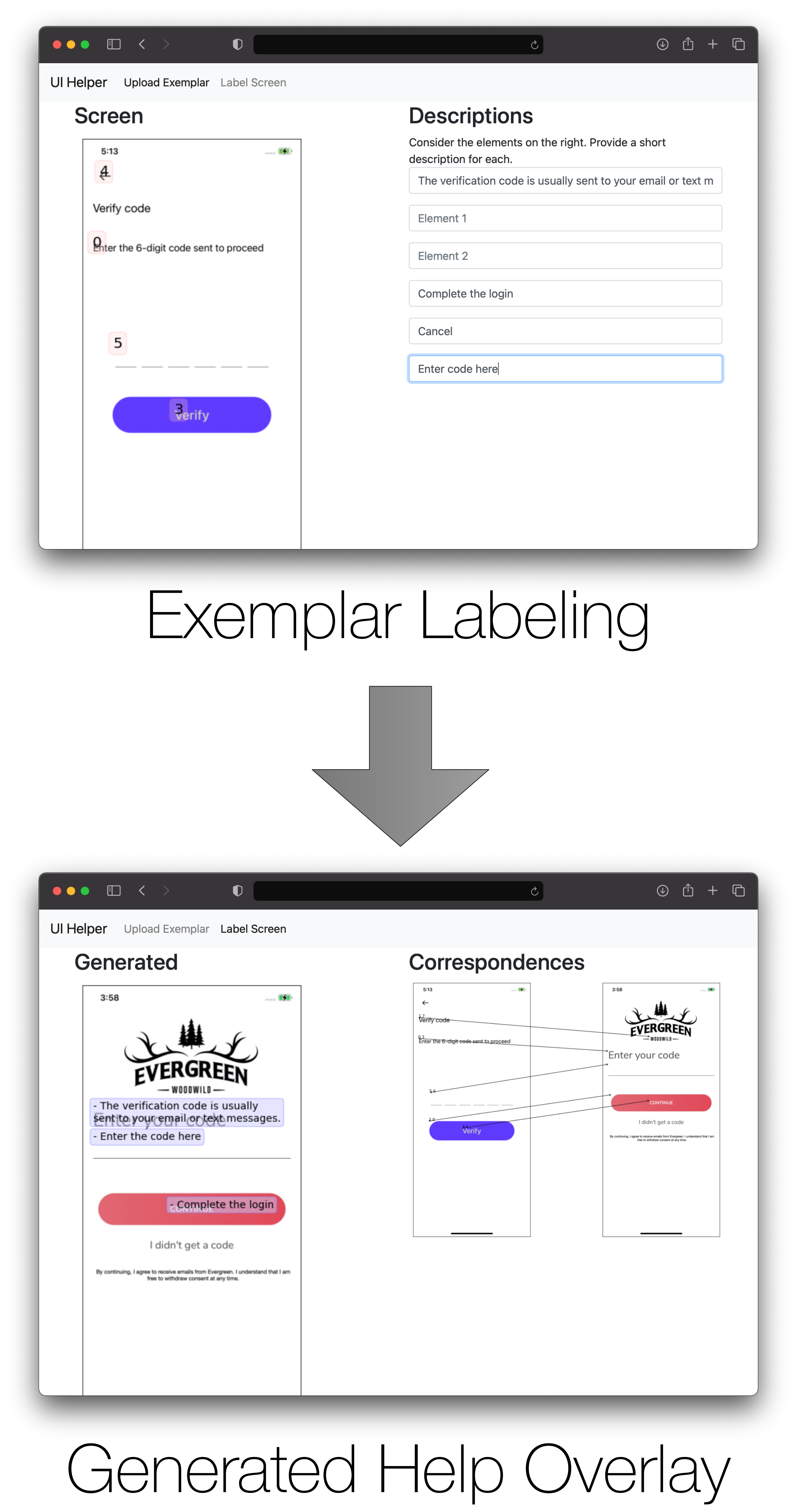}
    \caption{Coach marks are useful for uncovering functionality in apps. High-quality natural language descriptions of UI components can be difficult to generate, so we curated a small number of labeled examples from different app categories. Element descriptions from this labeled set are transferred onto unseen app screens of the same type using the correspondence mapping. \textcolor{black}{Depending on the use-case, the exemplar can be manually provided (\textit{e.g.,} developer wishes to label many similar screens at once) or automatically retrieved (\textit{e.g.,} a help-generation app uses a separate classifier to find an exemplar from a database of labeled screens.)}}
    \label{fig:my_exampleapp_coachmarks}
\end{figure}
We used our model to improve users' understanding of complex or newly installed apps by creating an infrastructure that could be used to crowdsource coach marks for apps.
Coach marks are instructional overlays that are sometimes shown to provide assistance to users when an app is first launched, and can be helpful for exposing UI functionality.
While it is possible to automatically generate natural language for describing screens \cite{wang2021screen2words} and widgets \cite{li2020widget} using deep models, they are often affected by surface-level appearance and may be prone to producing generic outputs \cite{li2020widget}.
\textcolor{black}{Building a model that produces natural language also introduces significant complexity that can be similar achieved with a correspondence mapping}.
A better approach might be to crowdsource users \cite{ramesh2011showmehow} or developers to write coach marks for screens in a subset of apps, and then apply our screen correspondence technology to map these coach marks onto a much larger set of screens with similar purposes. This idea builds on the template-based matching scheme of~\citet{yeh2011creating} for generating contextual help, and expands their idea beyond \textit{same-screen} applications to also \textit{intra-class} usage.

We applied our model's intra-class correspondence capabilities to automatically transfer annotations from one screen to another related app of the same category (Figure \ref{fig:my_exampleapp_coachmarks}).
We first populated a small database of instructional text for elements from app screens in one of the categories from our evaluation data. In a real implementation of this system, an interface would be created to allow users to author new instructional text for screenshots that they upload.
Each screen in the database was associated with its featurized elements as a key, and each instruction in the database was associated with its element.
Our current prototype is a proof-of-concept implementation where the user can upload a screenshot image file through a web interface.
On the uploaded screen, we perform a nearest-neighbor search to retrieve the screens in our database that are most similar.
If the distance is sufficiently close, we run our screen correspondence matching, which also returns a ``matching cost.''
If enough matches are discovered and the matching cost is below a heuristically set threshold, we directly render the annotations to the screenshot using image drawing APIs and display the annotated image.

In a complete implementation, the matching and rendering algorithms would be built into a mobile operating system and run on the user's mobile device so that it would not require the user to exit their current app to use our tool.
In the future, we plan to improve the user experience and investigate ways that these overlays could be surfaced contextually.


In this example, we show that accurate intra-class screen correspondence can facilitate transferring coach marks, which can help users discover new app functionality and documentation.
Other possible applications of screen correspondence to improving end-user usage include transferring more types of accessiblity meta-data, example-based re-targeting of UIs \cite{kumar2011bricolage} and using input redirection techniques to improve the accessibility of UI components \cite{zhang2017interaction}.
\subsection{UI Element Search Engine}
\begin{figure*}
    \centering
    \includegraphics[width=\textwidth]{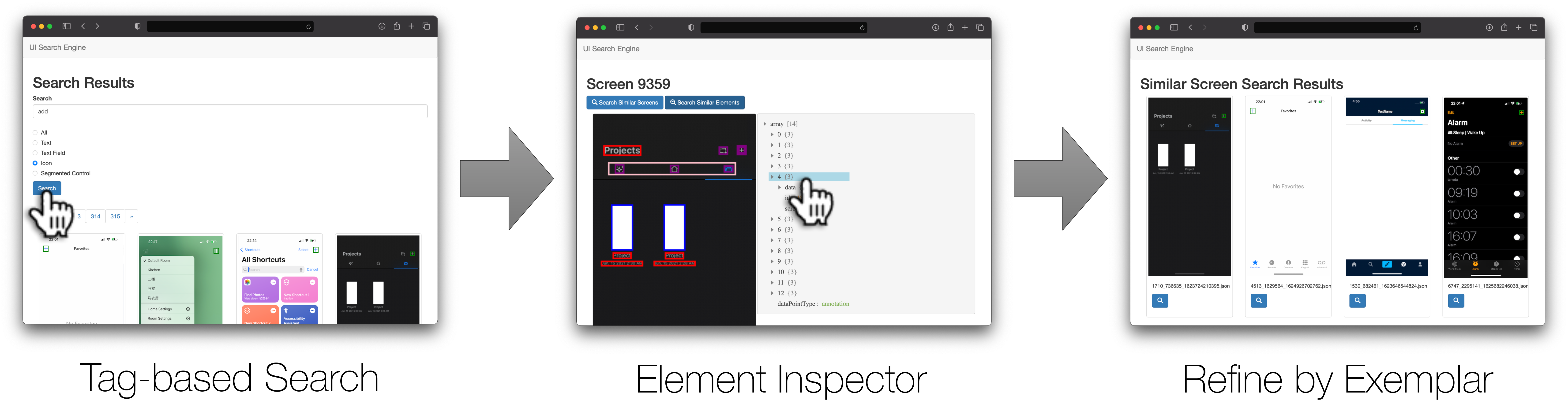}
    \caption{An example usage flow of our UI element search engine. The user first searches for icon elements that contain the ``add'' tag. The results page shows UI screens with a matching element highlighted (Left). The user selects a result screen where an add button is placed on the top right of the screen. The inspection page provides details about element info and allows searching for similar elements (Center). Another search query is run using the embedded element of interest. The new results are similar to the query in that they are all located at the top right of the screen and they appear to be used for adding items to a gallery (Right). This example shows how designers can start a search using natural language or tag-based queries then refine the results based on exemplars.}
    \label{fig:exampleapp_search}
\end{figure*}

UI search can help app designers find how concepts are expressed across apps and provide example starting points when designing a new app.
Previous work indexed databases of UI screens using visual properties \cite{bunian2021vins}, structural properties \cite{wu2021screen}, sketches \cite{huang2019swire}.
We focus specifically on returning relevant UI elements instead of screens, and leverage our model's \textit{intraclass} matching abilities to improve the search process.

We integrated our screen correspondence model into a UI search engine to support tag-based search and exemplar-based refinement.
The implementation of our UI search engine is a web app that indexed elements from 130,000 UI screens using a variety of metadata, including detected element classes, icon types, and text, which are stored in a database.
Our app features a search page, where users can first perform an initial search by entering text or tags in a search bar.
Results are returned based on matching attributes found in the property database.
Matching elements are shown in the context of their app screen and highlighted with a bounding box.
When a result is selected, users are brought to the element inspector page, where users can examine the properties of all elements on the screen.

One limitation of tag-based search is that it is difficult to specify target properties that do not belong to the pre-defined set of tags.
For example, a ``plus'' icon displayed on the top or bottom of a list may indicate adding \textit{to} the list while a ``plus'' icon displayed next to a list item is more likely to representing adding the item \textit{from} the list.
It would be difficult to disambiguate between these cases as they share the same tag.
Thus, we used our correspondence model to enable exemplar-based search refinement, which allows users to ``narrow in'' on more specific results.
To enable this functionality, we computed embeddings for UI elements in our database and stored this information into a vector data store which supports fast approximate nearest-neighbor search.
We added a ``search for similar items'' button on the element inspector page, which finds results with a high similarity to the target element according to the cosine similarity metric.
Figure \ref{fig:exampleapp_search} shows an example flow of our UI element search engine.


\subsection{Automated GUI Testing}
\begin{figure}
    \centering
    \includegraphics[width=20pc]{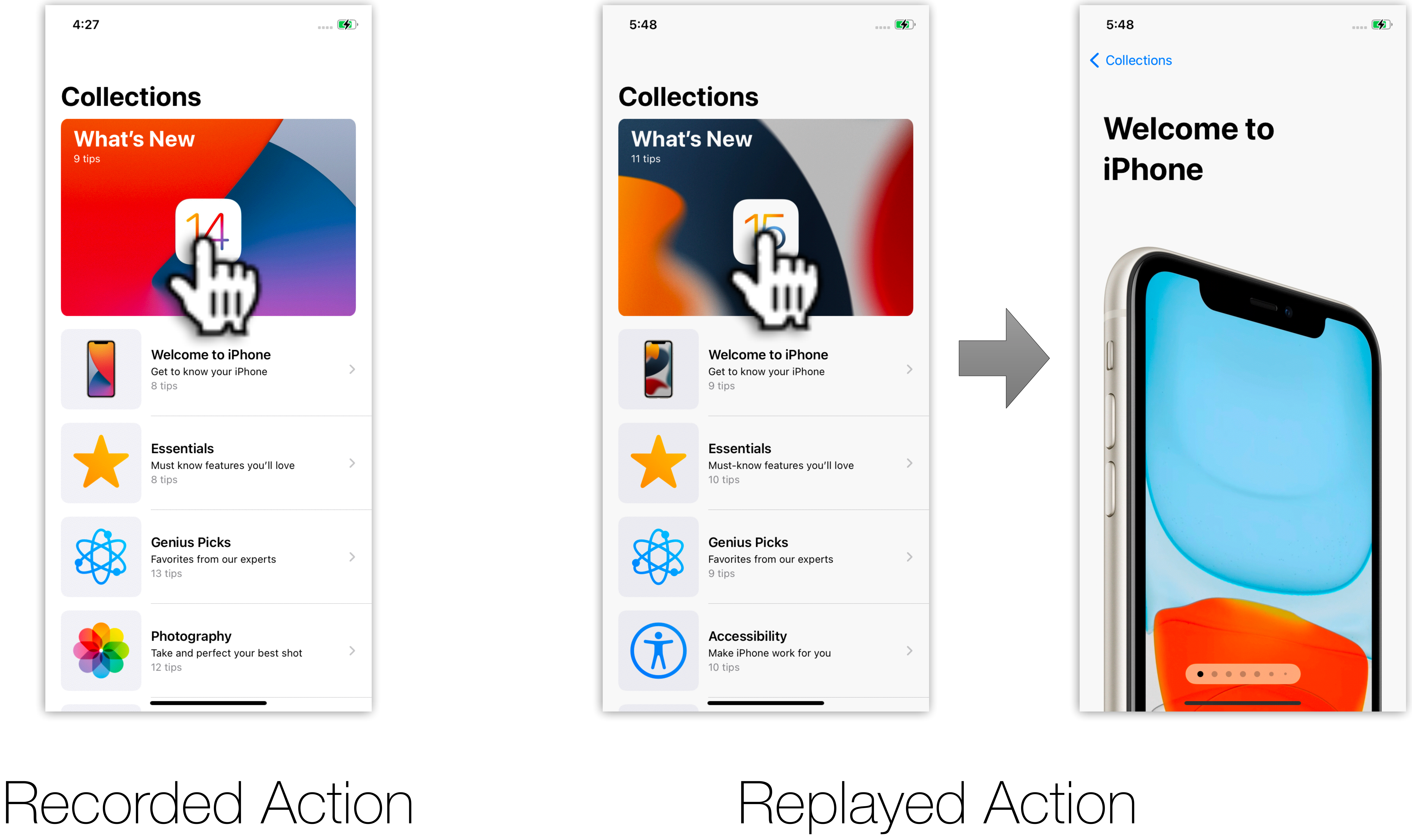}
    \caption{Automated UI testing techniques execute an interaction trace (either manually pre-defined or automatically generated) to detect functional regressions, visual regressions, and other unexpected behavior. Updated versions of apps may lead to small changes in layout and visual appearance and knowledge of same-screen correspondence can improve the consistency and robustness of tests. This example shows a automated application performing a previously recorded action, despite the target's appearance change.}
    \label{fig:exampleapp_guitest}
\end{figure}
Finally, we used our model to improve the robustness of automated GUI tests using our model's \textit{same-screen} matching capability.
Automated testing is useful for ensuring the quality of GUIs.
Specifically, visual-based methods can be employed in these systems to search for targets based on their rendered appearance, which allows for easier authoring of testing scripts and reduces the dependency of testing frameworks on specific UI toolkits \cite{chang2010gui}.
However, strong reliance on visual similarity may lead to failures caused by change in visual style, such as updated application theme or icons \cite{chang2010gui}.

In such applications, the quality of UI element matching is important for automated GUI testing because poor matching capability can lead to a failure to replicate recorded interaction traces in a scripted testing scenario, and repeated visits to the same screens in a random crawler stress test example.
As shown by previous work on screen similarity~\cite{disfani2022understanding}, methods that rely heavily on surface-level appearance may have high precision but low recall due to possible variations between UIs.
We applied our screen correspondence model to improve the robustness of these matches.

We built a prototype system that interacts with remotely connected smartphone devices through a VNC interface.
Our software sends commands through this interface to simulate interactions, such as clicking and swiping.
We also include a ``recording'' mode that allows a tester to record an interaction trace,
during which all of the screenshots and interactions are saved.
When replaying the interaction trace, the saved screenshots and interacted elements are used to match the current state of the VNC output.
Specifically, for each step in the saved trace, we identify the UI element with which the tester interacted, such as the button that was pressed.
Then, on the live VNC view, we find the corresponding element and apply the recorded interaction to it, similar to previous work on tutorial consumption \cite{zhong2021helpviz}.
Figure \ref{fig:exampleapp_guitest} illustrates how our automated tester navigates an app where the appearance of a target element has changed.
Used in conjunction with traditional template-matching techniques, which offer high precision but low recall, correspondence matching can help improve the overall performance of automated testers.


\section{Limitations \& Future Work}
\textcolor{black}{Our evaluation shows that correspondences can be automatically identified through machine learning and matching approaches. Some types of screens are more likely to have correspondences detectable by our system (\textit{e.g.,} Website Views and In-App purchases) than others (\textit{e.g.,} media players). The required accuracy level depends largely on the final application, since different use-case since different performance attributes.}
For example, using correspondences to generate contextual help (instructional overlays) may result in a better experience if only very confident matches are used, as incorrect instructions can lead to confusion and frustration from the user.
GUI testing and crawling is less tolerant to mistakes, since an incorrect action can make it impossible to access the rest of an application.
On the other hand, UI design search is more forgiving, since it can provide value if most of the returned elements are correct (does not need to be the top choice).
Our current evaluation does not account for the requirements of down-stream applications, although based on the example applications we implemented, we found them to provide acceptable performance.
We plan to further evaluate our system in down-stream applications.

A limitation of our current experiments is that they focus only on mobile UIs that belong to a set of 9 categories that we identified.
These 9 categories do not cover all possibilities of app screens, but they cover a considerable subset.
Our model is likely to perform better for complex app screens if given a small amount of annotations to fine-tune on.
Moreover, since we only use pixel information as input to our model, we believe that our approach is likely to generalize well to other types of graphical UIs that also represent their output as pixels.
In the future, we aim to replicate our experiments on other types of graphical UIs with varying screen sizes and shapes.

We see several opportunities to improve the performance of our system.
\textcolor{black}{Since our system relies on several individual components, it may be useful to quantify the performance of each separately. We used a pre-trained element detector model that produced noisy output for the correspondence matching. Previous work \cite{wu2021screen} has shown that element detectors perform poorly on more complex screens due to the increased number of elements and sometimes miss smaller elements. Future work could investigate a screen correspondence system that uses a more accurate element detector model or accepts manual annotations as input.}
More advanced matching techniques can also be employed, such those that consider multi-scale correspondence, which first process smaller sub-regions before merging their predictions globally.
Separately, prior work on image correspondence \cite{jiang2021cotr} has shown improved performance by scaling images during training and inference.
A similar idea could be applied to UIs by first predicting their UI hierarchy \cite{wu2021screen} and generating mappings for groups of elements.
Our model could also use different unsupervised pre-training objectives to help it build better representations of UI elements for our matching task \cite{Kocijan2019ASR,shen2021unsupervised}.

Our work focuses on mapping interchangeable elements with similar functionality between UI screens, however there are other relationships that can be modeled.
Categorization of different relations in language analogies \cite{lu2019emergence,gladkova2016analogy} show that antonym, categorical, and functional connections can enrich the expressiveness of language and rhetoric.
We plan to focus future modeling efforts on identifying and inferring a wider range of similar relationships that exist in UIs.

Finally, our work explores inferring UI functionality from a single previously encountered example, yet we believe our approach may extend to multiple examples \cite{fu2013cluster}.
For example, non-parametric machine learning methods such as the k-nearest neighbors algorithm often benefit from considering more than one example at a time.

\section{Conclusion}
In this paper, we explore \textit{screen correspondence} as a machine learning technique for inferring UI functionality by directly leveraging previously encountered examples.
We describe our model architecture and training procedure that incorporates information about UI semantics, appearance, and text when generating correspondence mappings between screenshots.
In a comprehensive evaluation with strong baselines, we show that our approach outperforms correspondence algorithms by leveraging multiple information sources found in UIs.
Finally, we show how three example applications of screen correspondence: \textit{(i)} transferring coach marks from related apps, \textit{(ii)} UI element search, and \textit{(iii)} automated GUI testing. Broadly, our work demonstrates the feasibility of learning UI semantics by mapping to prior examples.
\appendix
\section{Model Hyperparameters}
\begin{table}[!htb]
\begin{tabular}{lll}
\hline
Model         & Hyperparameter & Value \\ \hline
Screen Transformer & optimizer      & Adam  \\
              & learning rate             & 1e-4  \\
              & weight decay   & 1e-5  \\
              & dropout        & 0.25  \\
              & hidden size    & 256   \\
              & num layers  & 4     \\
              & num heads  & 4     
\end{tabular}
\end{table}

We trained our models with early stopping and stopped training when validation loss did not improve for 10 epochs.
We implemented our model using the PyTorch \cite{paszke2019pytorch} and PyTorch Lightning \cite{falcon2019pytorch} frameworks.

\section{UI Category Criteria}
We collected a small dataset of 9 screen categories for evaluation of our model's \textit{intra-class} correspondence capabilities.
We used the following guidelines to categorize apps.
\begin{itemize}
    \item \textit{Media Player} - A screen that allows users to play media content such as music or video. Usually contains controls for adjusting playback, volume, and sharing.
    \item \textit{In-App Purchase} - A screen that asks users to make a purchase for a subscription or to access some part of an app. Usually contains buttons for making the purchase, dismissing the screen, or signing up for a trial.
    \item \textit{Login} - A screen that asks users to log into an app or service. It may contain fields for entering username and password or buttons for third party authentication services.
    \item \textit{Permission Request} - A screen that asks users to enable some permission, which are usually associated with security settings such as location or camera access.
    \item \textit{Register} - A screen that asks the user to create an account. May contain a form to register or buttons for third party authentication providers.
    \item \textit{Pre-Login} - A screen that contains controls to access other parts of the app either by logging in or registering for an account. This usually comes before the login page.
    \item \textit{Pop-up} - A screen with a pop-up or dialog model that is displayed over other app content. Pop-ups may contain controls for accepting or dismissing it. For pop-ups that ask for permission or purchases, see other categories.
    \item \textit{Search} - A screen for entering and submitting a search query. May include a search bar and filtering controls.
    \item \textit{Website View} - A screen where an app opens an external website. May contain a URL bar and forward/backward controls.
\end{itemize}
\bibliographystyle{ACM-Reference-Format}
\bibliography{sample-base}

\appendix

\end{document}